\documentclass[twocolumn,showpacs,floatfix,prb]{revtex4-1}
\usepackage{graphicx}
\usepackage{float}
\usepackage{dcolumn}
\usepackage{bm}
\usepackage{amssymb}
\usepackage{amsmath}
\usepackage{hyperref} 
\hypersetup{bookmarks=true,unicode=true,colorlinks=true, urlcolor=blue,linkcolor=blue,citecolor=blue,filecolor=blue}

\begin{document}

\title{Band gap in $\bf{Bi_2Se_3}$ topological insulator nanowires:  magnetic and geometrical effects}

\author{P. Iorio, C. A. Perroni, and V. Cataudella}
\affiliation{CNR-SPIN and Physics Department "Ettore Pancini", \\ Universit\'a degli Studi di Napoli "Federico II", \\
Complesso Universitario Monte S. Angelo, \\Via Cintia, I-80126, Napoli, Italy.  }

\begin{abstract}
Stimulated by the recent realization of three dimensional topological insulator nanowire interferometers, a theoretical analysis of quantum interference effects on the low energy spectrum of $Bi_2Se_3$  nanowires  is presented. The electronic properties are analyzed in nanowires with circular, square and rectangular cross-sections starting from a continuum three dimensional model with particular emphasis on magnetic and geometrical effects. The theoretical study is based on numerically exact diagonalizations of the discretized model for all the geometries. In the case of the cylindrical wire,  an approximate analytical solution of the continuum model is also discussed. Although a magnetic field corresponding to half quantum flux is expected to close the band gap induced by Berry phase, in all the studied geometries with finite area cross-sections, the gap closes for magnetic fields typically larger than those expected. Furthermore, unexpectedly, due to geometrical quantum interference effects, for a rectangular wire with a sufficiently large aspect ratio and smaller side ranging from $50 \AA$ and $100 \AA$, the gap closes for a specific finite area cross-section without the application of a magnetic field. 
\end{abstract}

\maketitle

\section{Introduction}
Topological insulators (TI) are two \cite{Kane}  or three \cite{Fu} dimensional materials which, due to the bulk-boundary correspondence,  show new interesting features when confined in one direction and interfaced with the vacuum. Indeed, in the bulk gap,  they exhibit conductive surface states  protected by time reversal symmetry \cite{X-L,Hasan}.  These energy states connect the bulk valence band with the bulk conduction band through the formation of  Dirac cones \cite{Book1,Book2}. In this paper, we will consider Bismuth-based three-dimensional TI \cite{D.Hsieh1,D.Hsieh2,D.Hsieh3}, focusing on the binary compound $Bi_2Se_3$. Indeed, this system has a simple electronic structure: a single Dirac cone present at the $\Gamma$ point of the Brillouin zone and a bulk gap of the order of 0.5 eV \cite{D.Hsieh1,Nature}. 

In the slab configuration, when the two interfaces with the vacuum are sufficiently far from each other,  the structure of the Dirac cone is preserved. Close to the $\Gamma$ point,  due to the spin-momentum locking  at the surface, a rotation of the spin occurs when one considers different momenta in the Brillouin zone. The spin-momentum locking can be described by a Rashba model similar to that used for the description of spin-orbit couplings in two dimensional electron gases \cite{Rashba,Marigliano1}. However, if the thickness of the material consists of a few unit cells in the direction orthogonal to the interface, the electronic states of the two surfaces tend to hybridize through the thin bulk, hampering the formation of the Dirac cone \cite{Book1}. As a result, at the  $\Gamma$ point,  there is the opening of a gap whose size is typically smaller than that of the bulk gap. 
   
A different situation takes place when the TI is confined  into two directions as in the case of a nanowire. In a naive way, one can think that the surface states always communicate with each other favoring quantum interference effects. As a consequence, the opening of a small gap at the $\Gamma$ point occurs even for large nanowire cross-sections.  
In the nanowire, the opening of the gap is ascribed to the role played by the Berry phase, the geometrical phase which  characterizes the system relating the states on the cross-section perimeter with consequent real space spin-texture. Actually, the surface states not only are conductive, but they also show a particular spin rotation along the perimeter of the wire cross-section. Some of the mentioned properties have recently been reported. In presence of a magnetic field  Aharonov-Bohm oscillations have been detected showing that the transport is confined to the surface\cite{Peg}. Furthermore, in presence of a weak magnetic field applied parallel to the axis of the nanowire, experimental studies report peaks of the conductance at magnetic fields corresponding to fluxes close to half quantum flux ($\pm h/2e$) \cite{Nano}. The presence of this peak is explained by the fact that the effects of the Berry phase on the electronic spectrum are eliminated by the interference effects due to the magnetic field that is able to restore the Dirac cone. It is worth noticing that, in some cases,  the experimental value for the magnetic field able to close the Berry phase gap can also be $20\%$ larger than that corresponding to  half quantum flux \cite{Dufouleur}.
 
Stimulated by these results, in this paper, we study how the effects of the Berry phase in a $Bi_2Se_3$ nanowire open a gap at the Dirac cone for different geometries of the cross-section. We point out that the starting point of our analysis is not given by a surface model \cite{J.H.,Yi,Mariglia2,Bercioux}, but by a three dimensional  continuum bulk model considering not only circular \cite{Kundu,Imura}, but also square/rectangular cross sections for the nanowires.    
Through an exact numerical diagonalization of the discretized model, we study the electronic states of the nanowire with cylindrical, square and rectangular cross-sections since exact analytical solutions for the entire nanowires are not available. We find that the gap energy and the related behavior in the presence of magnetic field are qualitatively similar for different cross-sections but they quantitatively depend on the size and the geometry of the nanowire. 

In the case of cylindrical nanowires, we resort also to an approximated analytical solution valid in the continuum limit. Actually, a simple approximated analytic calculation, exact only close to the asymptotic limit of infinite radius, has been proposed in the case of a full cylinder \cite{Imura}.  In this work, we improve the approximations used in the previous paper, reporting an analytical solution which is valid for large but finite radii of the cylindrical nanowire. The effects of the Berry phase continue to be robust,  but, for finite radii, the closure of the gap takes place for magnetic fields larger than those corresponding to half quantum flux. In fact, we find that the additional magnetic field decreases with increasing the nanowire radius vanishing only in the limite of infinite radius. Concerning the numerical diagonalization, we use a grid with a variable mesh for the cross-section  in order to accurately simulate a circle of fixed radius. For radii larger than $100 \AA$,  the numeric approach perfectly reproduces the analytical results. With decreasing the radius, the numerical solutions show some deviations from the analytical results. 

We numerically analyze the electronic states of a nanowire with square/rectangular cross-sections considering not only different confining directions, but also different aspect ratios between the side lengths of the rectangle. In particular, in the case of a square cross-section, the electron density of the surface states exhibits peaks at the corners due to the boundary conditions and not to additional corner potentials as discussed in the literature \cite{Sen}. 

In all the geometries, only for infinite area cross-sections,  the closure of the gap in the presence of the magnetic field occurs for half quantum flux.  With decreasing the cross-section area, the gap closes for magnetic fields larger than those corresponding to half quantum flux showing a dependence on the size and geometry of the nanowire. Moreover, quite unexpectedly, we find a non monotonic behavior of the energy gap as a function of the cross-section characteristic length.  For all the geometries, we point out the presence of a minimum and maximum of the energy gap at the $\Gamma$ point in the range from $50$ to $100 \AA$.  As a consequence, close to the minimum, the magnetic field for the gap closing gets significantly reduced. In particular, in the case of a rectangle with aspect ratio equal to $1/6$,  the minimum coincides nearly with zero energy. This implies that the gap can be closed at a finite length without any magnetic field.  With decreasing the aspect ratio, close to the minimum of the gap, the system behaves decoupling the two sides with the shortest length along the perimeter, restoring in a certain sense  the case of a single confinement in the direction orthogonal to these sides. 

The paper is organized as follows. In Sec. II, the low energy continuum model for $Bi_2Se_3$ close to the $\Gamma$ point is considered; in Sec. III, the analytical and numerical solutions for the electronic states of the cylindrical nanowire are discussed; in Sec. IV, square/rectangular nanowires are investigated; in Sec.VI, conclusions and discussions.

\section{Low energy continuum model close to $\Gamma$ point}

$Bi_2Se_3$  is a material with a layered structure where each layer contains both atomic species forming a triangular lattice. The inversion of band responsible for the topological properties is due to the strong spin-orbit coupling of the $Bi$ atoms. 
Only  $p_z$ orbitals (perpendicular to the layers) of both $Bi$ and $Se$ atoms are relevant for the electronic structure near the Fermi level. The features of bulk and surface electronic states have been theoretically studied  and experimentally investigated by angle resolved photoemission spectroscopy (ARPES) \cite{Hasan,Wen}. At the $\Gamma$ point of the Brillouin zone, the TI $Bi_2Se_3$ has surface states corresponding to a single Dirac cone. 

As discussed in Ref. [\onlinecite{Nature}], an adequate description of the low energy bulk states around the $\Gamma$ point can be obtained if we consider only the states near the Fermi energy. Neglecting the contribution of $s$ orbitals, we can consider the combination of the $p$ orbitals of the two atomic species in a single unit cell according to their parity. The states relevant to our model are labelled as $|P1^{\pm}_z\rangle$ and $|P2^{\pm}_z\rangle$, where $\pm$ denote the parity. The important symmetries of the system are the time reversal  $\mathcal{T}$, the rotation symmetry $\mathcal{C}_3$ along z direction and inversion symmetry $\mathcal{I}$. In the  four-dimensional basis  ($|P1_z^+,\uparrow \rangle$, $|P2_z^-,\uparrow \rangle$, $|P1^+_z, \downarrow \rangle$, $|P2_z^-, \downarrow \rangle$), the representation of the cited symmetry operators is given by $\mathcal{T}= \mathcal{K} \cdot i\sigma^y \otimes \mathbb{I}_2$, $\mathcal{C}_3 = exp(i (\pi/3) \sigma^z \otimes \mathbb{I}_2 )$, and $\mathcal{I}=\mathbb{I}_2 \otimes \tau_z$ where $\mathcal{K}$ is the complex conjugation operator, $\sigma^{x ,y ,z}$ and $\tau_{x ,y ,z}$ denote the Pauli matrices in the spin and orbital space, respectively. In the presence of a magnetic field $B_z$ parallel to the z axis and neglecting non linear terms, the low energy continuum model  describing the electronic properties of $Bi_2Se_3$ close to the $\Gamma$ point is given by the following Dirac hamiltonian H:

\begin{eqnarray}
\label{eq:1}
H &= & M  \, \mathbb{I}_2 \otimes \, \tau_z -i  B \,  \sigma_z \otimes \tau_x  \partial_z \nonumber \\\
& - & iA \left[ \sigma_x \otimes \tau_x  \partial_x  + \sigma_y \otimes \tau_x (\partial_y+iB_zx) \right],
\end{eqnarray}
where $M$ is a mass operator which can be written as $M=M_0-M_2\nabla^2$, and the parameters $A$ and $B$ control the inter-orbital and inter-spin couplings. We choose a gauge with the vector potential equal to  $ \mathbf{A}=(0,B_zx,0)$.

For the three dimensional bulk with zero magnetic field, one gets from Eq. (\ref{eq:1}) the following $H_{3D}$ matrix at the momentum $\mathbf{k}=(k_x,k_y,k_z)$:

\begin{equation}
H_{3D}(\mathbf{k})=
\begin{bmatrix} 
\label{eq:2}
  M          &   B \, k_z      &       0               &    A \; k_{-} \\ 
  B \, k_z &   -M              &     A \, k_{-}      &         0       \\
  0            &    A\, k_{+}   &       M              &         -B\, k_z\\
  A \, k_+   &     0            &        -B\,k_z      &          -M
\end{bmatrix},
\end{equation}
with $k_+=k_x+ik_y$ and $k_-=k_x-ik_y$. 
The topological nature of the material is determined by the relative sign between $M_0$ and $M_2$. If the ratio between the two parameters is positive, then the insulating state is $\mathbb{Z}_2$ with topological index $\nu=0$  i.e. the material is a trivial insulator. Otherwise, one gets $\nu=1$ and the material is a strong topological insulator. The strong topological insulator has surface states protected by strong and weak disorder unlike the weak topological insulator that in presence of disorder pushes the metal surface state into the bulk, opening a gap in the Dirac cone\cite{Zohar}.

If we look at the problem of a slab of this material\cite{Wen}, new surface states appear in the gap energy around zero energy.  For example, if we confine along z direction and the slab thickness is very large, the positive surface energy is given by $E_1(k_x,k_y)=A \sqrt{k_x^2+k_y^2}$. On the other hand, if we confine along x direction (with slab thickness still very large), one gets  the positive surface energy $E_2(k_y,k_z)=\sqrt{A^2k_y^2+B^2k_z^2}$.  The difference of the two energies $E_1(k_x,k_y)$ and $E_2(k_y,k_z)$ is related to the slope of the Dirac cone in a two dimensional  Brillouin zone which depends on the values of the parameters $A$ and $B$ in Eq. (\ref{eq:1}). 

\begin{figure}[h!]
\centering
\includegraphics[width=8.5cm]{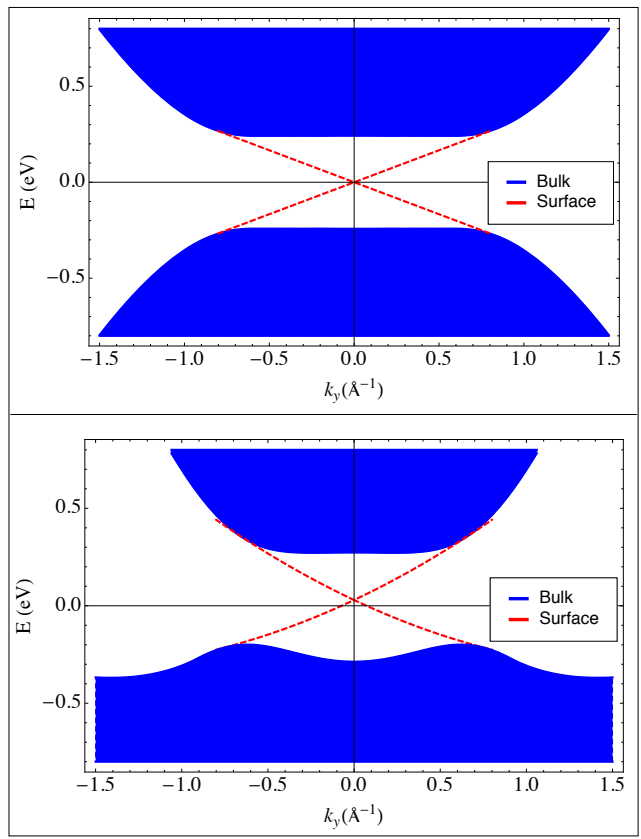}
\caption{\label{fig:1} The Energy spectrum $E(k_x=0,k_y)$ in eV ($k_y$ (in unit of $\AA^{-1}$). The projected bulk energies are indicated by the blue areas, the surface energies due to the confinement along z direction by red dashed lines.
Top panel: energies with the choice of parameters made in this paper ($M_0=-0.28 \,eV$, $M_2=40.00 \, eV \AA^2$ and $A=B=3.33\, eV \AA$).
Bottom panel: energies with the fit parameters taken from the \textit{ab initio} calculations in Ref.[\onlinecite{Nature}]. }
 \end{figure}
 
The tunable parameters in the Hamiltonian can be obtained by fitting the band structure obtained by \textit{ab initio} calculation \cite{Nature,Wei} close to $\Gamma$ point.  In this paper, we have further reduced  the number of free parameters choosing $A=B$ and assuming hole-particle symmetry.  However, as shown in Fig. \ref{fig:1}, the restriction is not severe.  In the rest of the paper, we fix:  $M_0=-0.28 \,eV$, $M_2=40.00 \, eV \AA^2$ and $A=B=3.33\, eV \AA$.

In the following, as discussed in the introduction, we accurately investigate the electronic properties of the surface and bulk states for nanowires with different cross-section geometries. Due to the electron-hole symmetry, we typically focus on the electronic spectrum corresponding to positive energies.  

\section{Cylindrical nanowire: analytical and numerical solution}
In the case of a wire with a circular cross-section and with translation invariance along z axis, one can exploit the symmetries of the geometry to obtain an approximate analytical solution. A solution of the problem, that is exact in the asymptotic limit of infinite radius, has been proposed in Ref.[\onlinecite{Imura}]. In fact, this solution is based on an adiabatic  separation between the radial direction (``fast variable") and the angular variable (``slow variable"). In this section, we improve that analytical approach introducing a better description of the radial component of the wavefunction. Moreover, we will test the analytical approximated solution by using exact diagonalization of a discretized model. 

Following the approach proposed in Ref.[\onlinecite{Imura}], it can be shown that the approximate surface eigenfunctions of the hamiltonian in Eq. (\ref{eq:1}) can be found in the form

\begin{equation}
\label{eq:3}
\psi_m (k_z;r,\phi)= \left[\frac{1}{\sqrt{2}}\mathbf{u}_1(\phi)+ c^\pm(m,k_z)\mathbf{u}_2(\phi)\right] f(r) e^{im\phi},
\end{equation}
where polar coordinates ($r,\phi$) are used. In Eq. (\ref{eq:3}), we have introduced the two orthogonal four component vectors

\begin{equation}
\mathbf{u_1}(\phi)=
\begin{bmatrix} 
   \, 1 \, \\ 
   \, i    \,     \\
   \, e^{i\phi}  \,     \\
   \, ie^{i\phi}   \,
\end{bmatrix},
\quad
\mathbf{u_2}(\phi)=
\begin{bmatrix} 
\label{eq:7}
   \, 1 \, \\ 
   \,- i    \,     \\
   \, -e^{i\phi}  \,     \\
   \, ie^{i\phi}   \,
\end{bmatrix},
\end{equation}

where $f(r)$ is a solution of the following differential equation:

\begin{equation}
\label{eq:4}
M_0f(r)-M_2 \left(\frac{f'(r)}{r}+f''(r)\right) = -A f'(r).
\end{equation}

If one defines  $a=M_2M_0/A^2$ and $z=Ar/M_2$, the solution of (\ref{eq:4}) with boundary condition $f(R)=0$ ($R$ is the cylinder radius) can be found in terms of combinations of hypergeometric functions and Laguerre polynomials \cite{Slater,kummer1,Arfken}:

\begin{equation}
f(z)=e^{\frac{1}{2} \left(1-\sqrt{1+4 a}\right) z}\frac{H_y(a,R)L_a(a,z)-H_y(a,z)L_a(a,R)}{H_y(a,R)},
\end{equation}
where
\[
H_y(a,z)=\text{HypergeometricU}\left[ -\frac{1-\sqrt{1+4 a}}{2 \sqrt{1+4 a}},1,\sqrt{1+4a} z \right]
\]
and
\[
L_a(a,z)=\text{LaguerreL}\left[\frac{1-\sqrt{1+4 a}}{2 \sqrt{1+4 a}},\sqrt{1+4 a} z \right].
\]

We note that the validity of this solution is extended to lower values of $r$ compared to what done in Ref.[\onlinecite{Imura}] since in this paper the term $(1/r)\partial_r$ of the Laplacian is taken into account.

In equation (\ref{eq:3}) the coefficient  $c^\pm(m,k_z)$ is given by:

\begin{eqnarray}
\label{eq:19}
c^\pm(m,k_z)&= &\pm\frac{1}{|E_{m,k_z}|}\biggl[ \biggl(\frac{M_2}{R^2}\langle \frac{R^2}{r^2}\rangle   -\frac{A}{R}\langle \frac{R}{r}\rangle\biggr) \times \nonumber\\\
&&\biggl(m+\frac{1}{2}- \frac{\Phi \langle r/R\rangle }{\Phi_0\langle R/r\rangle }\biggr) +iBk_z\biggr],
\end{eqnarray}
where the sign $\pm$ corresponds to positive and negative eigenvalues given by:
 
\begin{eqnarray}
\label{eq:8}
E_{m,k_z} &= & \pm\biggl[\left(\frac{M_2}{R^2}\langle \frac{R^2}{r^2}\rangle   -\frac{A}{R}\langle \frac{R}{r}\rangle  \right)^2\times\nonumber \\\
 &&\biggl(m+\frac{1}{2}  - \frac{\Phi \langle r/R\rangle }{\Phi_0\langle R/r\rangle } \biggr)^2+B^2k_z^2\biggr]^{1/2}.
\end{eqnarray}

In Eq. (\ref{eq:8}), the symbol $\langle \cdot\rangle $ indicates the average value over the radial function $f(r)$, $\Phi$ is the flux threaded by the cross-section, and $\Phi_0=h/(2e)$ is the magnetic quantum flux.

\begin{figure}[h!]
\centering
\includegraphics[width=8.5cm]{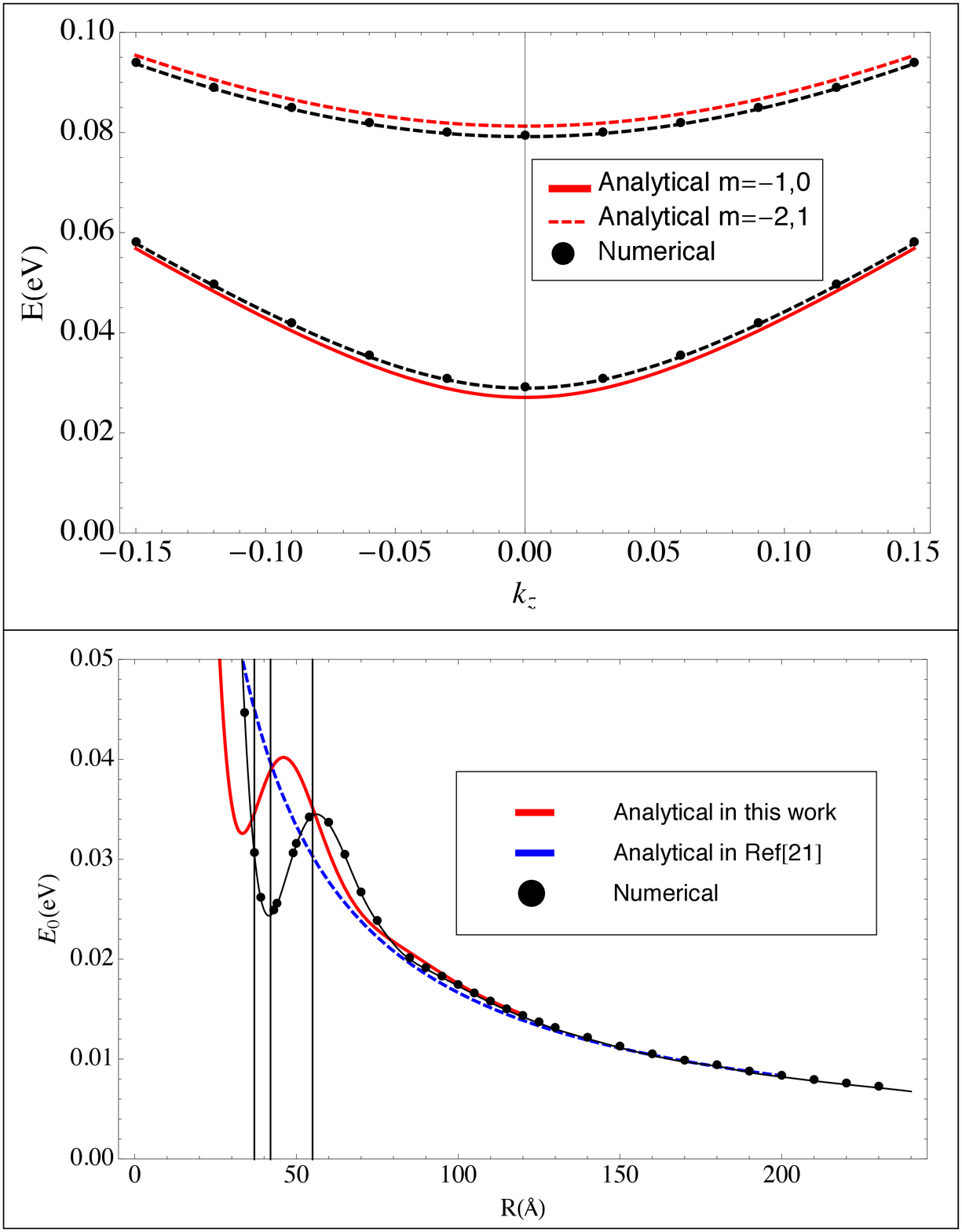}
\caption{Top panel: energy dispersion (in units of $eV$) for the first and second sub-band with positive energy as a function of $k_z$ (in units of $A/M_2 \simeq 0.083 \AA^{-1}$) for $R=65 \AA$. Energies come out from two approaches: Eq. (\ref{eq:8}) of our approximate analytical solution (red lines) and those derived from the numerically exact solution (black lines with circles). Bottom panel: gap energy (in units of $eV$) at $k_z=0$ as a function of the cylinder radius $R$ (in units of $\AA$) for three types of approaches: Eq. (\ref{eq:8}) of our approximate analytical solution (red line),  energy derived from the analytical solution of  Ref.[\onlinecite{Imura}], and energy obtained from the numerically exact solution (black line with circles).}
\label{fig:3} 
\end{figure}

Notice that in Eq. (\ref{eq:8}) the parameter $m$ labels the cylinder sub-bands whose spectrum can be also within the bulk gap. Moreover,  in the absence of the flux $\Phi$, thanks to the factor 1/2 close to the parameter $m$, all the sub-bands are doubly degenerate in $m$. In particular, the lowest sub-band is degenerate for $m=-1$ and $m=0$, therefore for this energy there are two associated eigenvectors $\psi^{\pm}_{-1}(k_z;r,\phi)$ and $\psi^{\pm}_0(k_z;r,\phi)$.
In the top panel of Fig. \ref{fig:3}, for $R=65 \AA$, we show the positive eigenvalues (red lines) associated to $m=-1,0$ (first sub-band) and $m=-2,1$ (second sub-band) as obtained by Eq. (\ref{eq:8})  as a function of $k_z$. As shown in the same panel, there is a good agreement between these analytical results with those (black lines with circles),  numerically exact,  obtained by using exact diagonalization of a discretized model (see Appendix for details about the numerical method).

As expected, a small gap of the order of $0.04eV$ (about one tenth of the bulk gap) is obtained. It is worth noticing that the opening of the gap has nothing to do with finite size effects, but it is directly related to the factor 1/2 close to the parameter $m$ in Eq. (\ref{eq:8}). This factor 1/2  represents the signature of the Berry phase. Actually, the Dirac cone present at 
$\Gamma$ point is removed by the Berry phase effect stemming from adiabatic paths around the cylinder perimeter \cite{Nano,J.H.}.  This result, that is reproduced here starting from a three dimensional bulk model (full cylinder), has been first discussed in a two dimensional cylindrical surface (hollow cylinder) since the key ingredient resides on the quantum interference effects on the surface \cite{Yi,Kundu}. However, when the cylinder radius gets reduced, the solutions are no longer strictly confined on the surface and the electrons experience a finite probability to stay into the bulk. 

We, then, focus our study on how the gap changes as a function of the cylinder radius $R$. In the bottom panel of Fig. \ref{fig:3}, we compare the gap computed with E q.(\ref{eq:8}), that calculated in Ref.[\onlinecite{Imura}], and that obtained numerically by using exact diagonalization. The three approaches coincide at large values of $R$ confirming the asymptotic validity of the approximated analytical solution, but they exhibit significant differences at small $R$.
However, even if quantitatively different in this regime, both our analytical solution in Eq. (\ref{eq:8})  and the numerical one  show a minimum-maximum structure for radii $R$ ranging from 40 to 60 \AA.  We remark that this behavior is not present in previous analyses proposed in the literature where only the limiting case of very large $R$ has been investigated. 
This non monotonic behavior of the gap energy is somehow unexpected and is one of the main results of this work. 

\begin{figure}[h!]
\centering
\includegraphics[width=8.5cm]{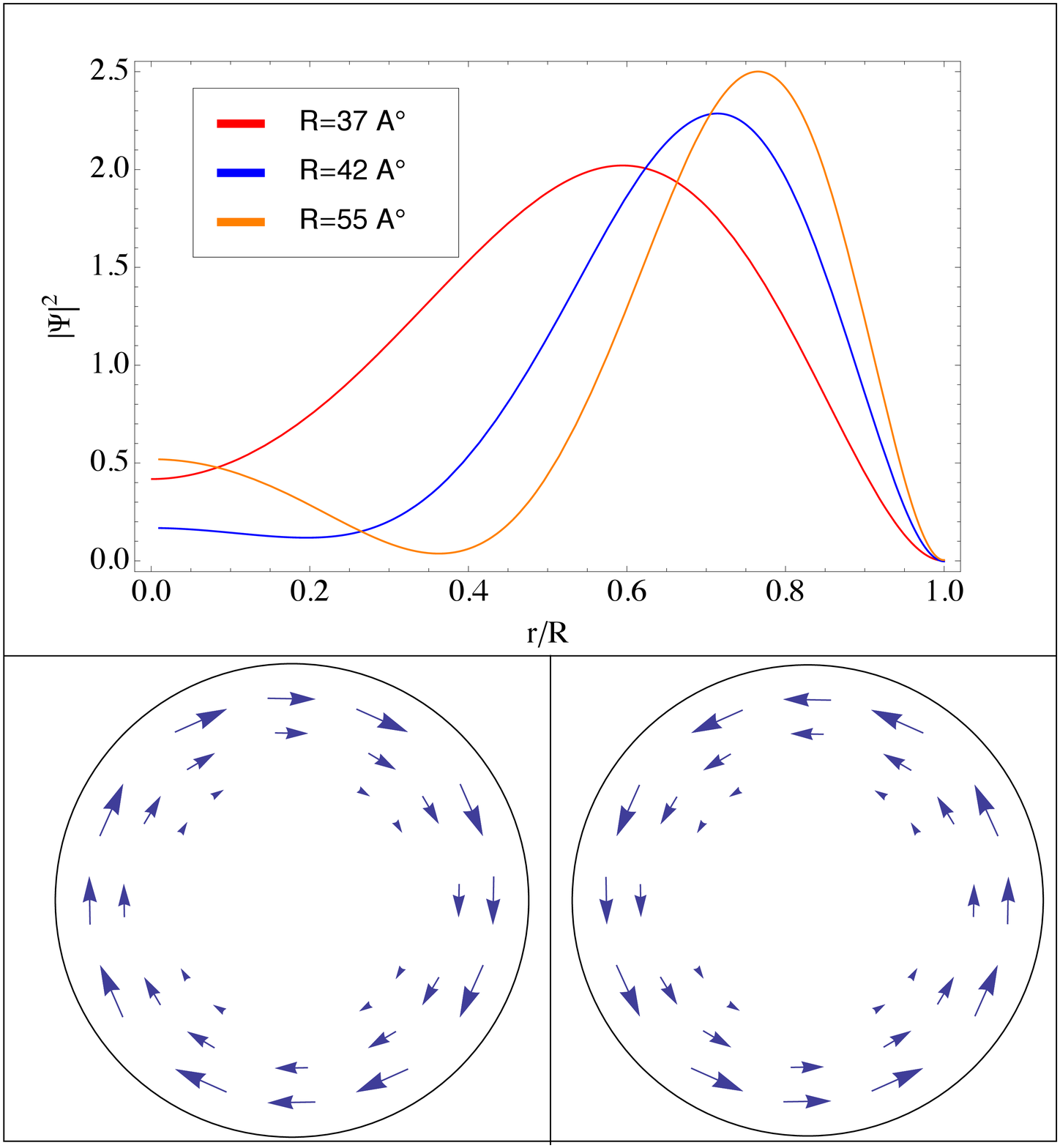}
\caption{\label{fig:4} Top panel: Radial probability density of the first sub-band state at $k_z=0$ as a function of the radial coordinate for different cylinder radii identified by the dotted vertical lines in the bottom panel of Fig. \ref{fig:3}. Bottom panel: spin-textures of the first sub-band state at $k_z=0$ obtained numerically for a cylinder of radius  $R=65 \AA$.  The modulus of the spin values is proportional to the length of the arrows.}
\end{figure}

We should emphasize, as already mentioned, that the observed gap has nothing to do with the bulk gap of Fig. \ref{fig:1} (of the order of $0.5$ eV). It is, as mentioned above, a consequence of the Berry phase associated to the electronic state peaked at the cylinder surface and can be traced back to the term $1/2$ present in Eq. (\ref{eq:8}).  However, the gap value is controlled in Eq. (\ref{eq:8}) by the factor $\left(\frac{M_2}{R^2}\langle \frac{R^2}{r^2}\rangle   -\frac{A}{R}\langle \frac{R}{r}\rangle  \right)^2$  that strongly depends on the radial wavefunction through the average values of $R/r$ and $R^2/r^2$. When $R$ is reduced and the radial wavefunction is not any longer strictly peaked at the cylinder surface, this factor provides an extra contribution to the gap. Naively one would have expected a simple  monotonic increase of the gap since $\langle R/r\rangle$ and $\langle R^2/r^2\rangle$ should increase when $R$ decreases. Our calculation, instead, shows a non monotonic behavior that is confirmed by both the approximate analytical solution and the ``numerically exact" solution. This behavior depends on how the ``surface state" fills up the cylinder upon reducing the radius $R$. In order to clarify this point, in the upper panel of Fig. \ref{fig:4}, we plot the first sub-band wavefunction probability density, obtained numerically, for the radii corresponding to the lowest analyzed value  ($37\AA$, red line) and for the $R$ values corresponding to the minimum ($42\AA$, blue line) and maximum ($55\AA$, orange line) (radii corresponding to vertical dotted lines in the bottom panel of  Fig. \ref{fig:3}).  We can see how, for the the radius corresponding to the maximum, the wave function extends to lower values of $r$ compared to the case $R=42\AA$, thereby entering the bulk more than the wave-function corresponding the minimum.

One of the main features of the TI is related to the spin texture. We start from the analytic solution to understand the complex spin behavior of these nanowires.  For simplicity, we consider the case with $k_z=0$. In the absence of magnetic field, as reported in Eq. (\ref{eq:8}), the two states for $m=0$ and $m=-1$ correspond to the same energy and they can be written as  
\begin{eqnarray}
\psi^{\pm}_{0}(r,\phi)  & = &\biggl(\frac{1}{\sqrt{2}}\mathbf{u_1(\phi)}+c^\pm(0)\mathbf{u_2(\phi)}\biggr)f(r),\nonumber\\
\psi^{\pm}_{-1}(r,\phi) &= &\biggl(\frac{1}{\sqrt{2}}\mathbf{u_1(\phi)}+c^\pm(-1)\mathbf{u_2(\phi)}\biggr)e^{-i\phi}f(r),
\label{eq:9}
\end{eqnarray}
where $c^\pm(m)$ is given in Eq. (\ref{eq:19}).  Then,  in the two dimensional subspace at fixed energy, we consider the matrix elements of the three spin components $S_x=\sigma_x \otimes \tau_z$, $S_y=\sigma_y \otimes \tau_z$, and $S_z=\sigma_z \otimes \mathbb{I}$ (written on the basis of  the Hamiltonian in Eq. (\ref{eq:1})). Considering only the lower positive sub-band, one gets a $2 \times 2$ matrix depending on $(r,\phi)$ for each spin component:
 \begin{equation}
\begin{pmatrix} 
\label{eq:10}
  \langle \psi^{+}_{0}(r,\phi) |S_i|\psi^{+}_{0}(r,\phi)  \rangle      &      \langle \psi^{+}_{0}(r,\phi) | S_i |  \psi^{+}_{-1}(r,\phi)  \rangle  \\ 
   \langle \psi^{+}_{-1}(r,\phi) |S_i|\psi^{+}_{0}(r,\phi)  \rangle      &      \langle \psi^{+}_{-1}(r,\phi) | S_i |  \psi^{+}_{-1}(r,\phi)  \rangle \\
  \end{pmatrix},
\end{equation}
with $i=x,y,z$. For the two in-plane spin components  ($i=x,y$), only off-diagonal matrix elements are different from zero, while, for the component along z, all the matrix elements vanish.
At this point, in order to find a pseudo average value of the spin, we diagonalize the matrix (\ref{eq:10}) for each spin component at fixed  $(r,\phi)$ and we plot the eigenvalues. The eigenvalues for $k_z=0$ of the $x$ component are: $\lambda_x =\pm  f(r) \cos(\phi)$; the two eigenvalues of the $y$ component are: $\lambda_y =\pm  f(r) \sin(\phi)$; the eigenvalues of $z$ component are clearly zero. 
Therefore, the spin eigenvalues are modulated by the radial part of the wave-function.   
For each spin component, the choice of the eigenvalue with varying $(r,\phi)$ is made by ensuring that the phase of the corresponding spin eigenvector is continuous. Therefore, one gets two main spin textures: clockwise, corresponding to spin eigenvalues proportional to the vector $f(r)\left( \cos(\phi), \sin(\phi),0 \right)$; counterclockwise, corresponding to spin eigenvalues proportional to the vector $- f(r)\left( \cos(\phi), \sin(\phi),0 \right)$.  The procedure described above can be easily extended at finite values of $k_z$.  

As discussed in Appendix, at fixed energy and $k_z$, the method of the diagonalization of spin components can be also used starting from the eigenvectors obtained numerically.  We report in the bottom panel of Fig. \ref{fig:4} what we have obtained in the case of the numerical eigenvectors at $k_z=0$.  Indeed, the spin rotates around the surface both clockwise and counterclockwise. The agreement with the spin textures calculated analytically is good. In particular, the modulus of the spin values (proportional to the length of the arrow) directly follows the behavior radial probability density (some examples are in the top panel of Fig. \ref{fig:4}).   

\begin{figure}[h!]
\centering
\includegraphics[width=9.0cm]{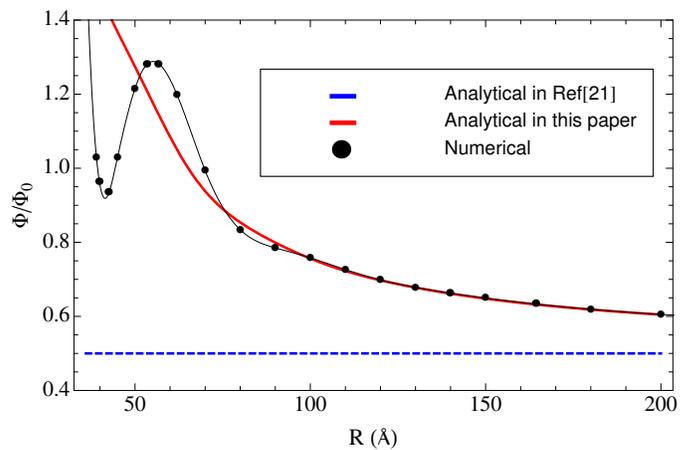}
\caption{\label{fig:5} Magnetic flux $\Phi$ (in units of quantum flux $\Phi_0$) corresponding to the gap closure as a function of the cylinder radius $R$ (in units of $\AA$).  Curves derive from three approaches: analytical solution of Ref. [\onlinecite{Imura}] (dash blue), improved analytical  solution (red) discussed in this paper and numerical method (black with solid circles).}
\end{figure}

The application of a magnetic field along the translational invariant z axis  of the wire introduces a magnetic flux $\Phi$ which can change the gap value as can be seen from Eq. (\ref{eq:8}). There are many studies \cite{Nano,J.H., Hao} about the effects of a magnetic field in TI nanowires showing the formation of a one dimensional  band for  electrons on the cylinder surface.  In the presence of a weak magnetic field applied parallel to the axis of the nanowire, experimental studies\cite{Nano} report peaks of the conductance at magnetic fields corresponding to fluxes close to half quantum flux.
 For very large cross sections, the specific values for which the gap closes correspond to semi integer quantum fluxes ($\Phi_0/2$, $3\Phi_0/2$,...) and they are distinguished from integer values found in ordinary  Aharonov-Bohm effects.  
The analytical result provided by Eq. (\ref{eq:8}) and confirmed by the numerical calculations shows that the gap closure depends on the size of the wire (see Fig. \ref{fig:5}). In fact, according to Eq. (\ref{eq:8}), the gap closes at $\frac{\Phi}{\Phi_0}=\frac{1}{2} \frac{ \langle R/r\rangle }{\langle r/R\rangle }$, a value strongly depending on the radial wavefuntion and, therefore, on the radius $R$ of the cylinder. The flux value closing the gap recovers $\Phi_0/2$ only in the asymptotic regime ($R\mapsto\infty$). Indeed,  only for large radii, the ratio 
$\frac{ \langle R/r\rangle }{\langle r/R\rangle }\mapsto \frac{1}{2}$. The extrapolation of the fit of our analytical curve to infinity confirms this result. On the other hand, for small radii, the gap does not close at half quantum flux. This is, then,  in contrast with what stated in the literature for the strictly two-dimensional case \cite{Peg,Nano,Yi} and what comes out including the magnetic field contribution in the approach of Ref.[\onlinecite{Imura}] (dashed line in Fig. \ref{fig:5}).

We point out that, for large radii, all the analytical and numerical approaches are completely consistent. However, as shown in Fig. \ref{fig:5}, for small radii, the numerical data depend on the radius again in a non monotonic way differing from those derived from the proposed analytical solutions. More precisely, the numerical results show the presence of a maximum and minimum in the magnetic flux corresponding to the gap closure.  The presence of these stationary points is found for values of the radii equal to those found for the non monotonic behavior of energy gap when the magnetic field is not applied. The fully numerical approach looks very accurate since the gap closing magnetic flux is expected to be sensitive to the behavior of the gap in the absence of the magnetic field.  We remark that even the improved analytical solution proposed in this paper does not predict the existence of the minimum and maximum for the magnetic flux.

\section{Rectangular nanowire}
In this section, we discuss the case of a wire with a rectangular cross-section. The problem will be addressed only numerically, since, unlike the cylinder, an approximate analytical solution is not available for a generic cross-section and orientation. Through a simple tight binding procedure, we analyze the surface states with their spin polarizations along the perimeter of the wire, for different confinements along the axes of the crystal lattice. This type of tight binding procedure does not consider the complex structure of the material since the triangular lattice is approximated by an effective square lattice \cite{Sen}. To get the discrete model, we always start from Eq.(\ref{eq:1}), which is discretized with a fixed lattice parameter for any direction of the axes.  All our results are obtained with a lattice constant $a=10 \AA$. However, we remark that, also for the smaller lattice parameter $a=5 \AA$, we find qualitatively similar results.  

As for the cylinder, we analyze the surface states as a function of the wire size. In particular, we study the case of a wire with square section in the case of out of plane translational invariance  (along z-axis as for the cylinder), but we also analyze the case of translational invariance in the y axis direction (due to the hamiltonian symmetries, the case invariant in the x direction is indistinguishable).  Also in this case the choice of parameters of the model is equal to that of the cylinder, therefore we consider $A=B=3.33 eV \AA$ in Eq. (\ref{eq:1}). In this way, as discussed in Section I, the spectrum for the two types of confinement remains the same. Therefore, for the wire with translational invariance in z or y direction, the same energy dispersion  is obtained and, consequently, also the same values of the magnetic flux for the gap closing. To analyze the size effects on the rectangular geometry, we investigate cross-sections with aspect ratio equal to 1/1, 1/2, 1/3, and, finally, 1/6.

\begin{figure}[h!]
\centering
\includegraphics[width=8.5cm]{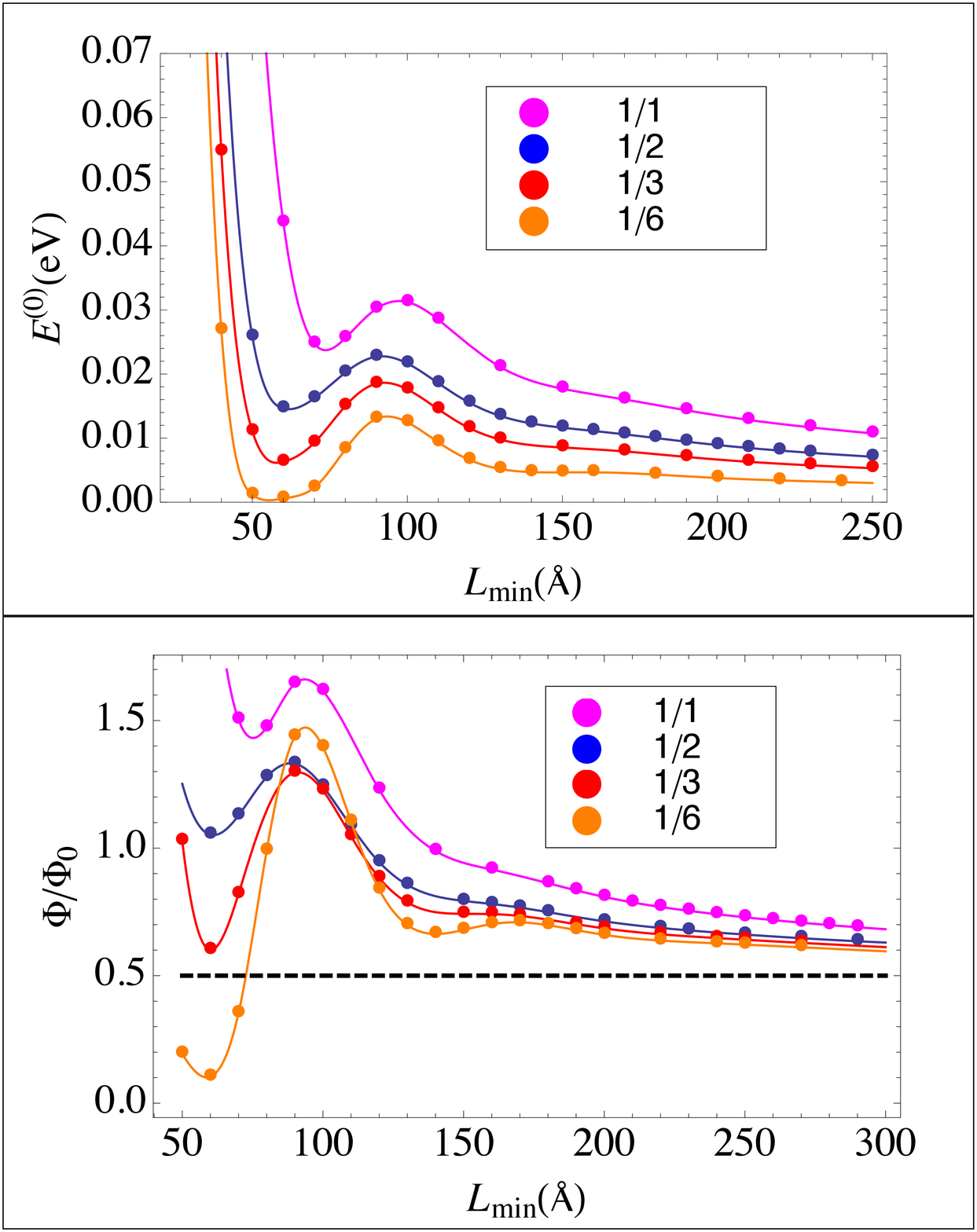}
\caption{\label{fig:7} Upper panel: gap energy (in units of eV) as a function of the length of smaller side $L_{min}$ (in units of $\AA$) in the case of rectangular cross-sections with different aspect ratios. Lower panel: magnetic flux (in units of quantum flux $\Phi_0$) corresponding to gap closing as a function of $L_{min}$ (in units of $\AA$) in the case of rectangular cross-sections with different aspect ratios.}
\end{figure}

In the upper panel of Fig. \ref{fig:7}, we plot the energy gap as a function of the length of the smaller side $L_{min}$.  As in the case of cylindrical geometry, the energy gap tends to zero in the limit of infinite length. A numerical fit of our results provides the following asymptotic behaviour: $E \sim \frac{c_1}{L}+\frac{c_2}{L^2}$ (where $c_1$ and $c_2$ are parameters of the fit and depend from the investigated aspect ratio), in strong analogy with what we have found for the cylindrical nanowire where we got $E \sim \frac{M2}{2R^2}-\frac{A}{2R}$.  Furthermore, also for this geometry, we find a non monotonic behavior with a maximum and a minimum for lengths between $50$ and $100 \AA$. In particular, the minimum in the energy gap gets smaller with reducing the aspect ratio. The case relative to the aspect ratio 1/6, that is the lowest we studied, is really interesting. In fact we see that, for   $L_{min}$ around $60 \AA$, the gap closes almost completely without the introduction of the magnetic field. As for the cylinder, we ascribe the presence of the minimum to geometrical effects, then, we find that specific wavefunction density probability due to purely geometrical effects is able to compensate topological effects due to the Berry phase. This is one of the main results of the present work. 

\begin{figure}[h!]
\centering
\includegraphics[width=8.5cm]{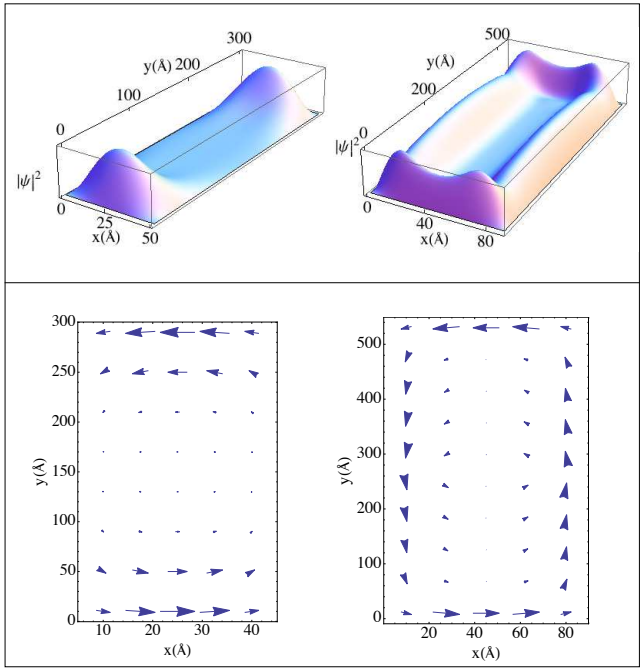}
\caption{\label{fig:8} Upper panel: spatial probability density of the first subband state at $k_z=0$ for a rectangular wire with aspect ratio equal to 1/6. Left plot for $L_{min}=60 \AA$, right plot for $L_{min}=90 \AA$. Lower panel: spin orientation corresponding to the first subband state at $k_z=0$ for a rectangular wire with aspect ratio equal to 1/6. Left plot for $L_{min}=60 \AA$, right plot for $L_{min}=90 \AA$. For simplicity, in the two plots, only one of the two possible spin orientations is shown.}
\end{figure}

Another important aspect that emerges from this analysis is related to the properties of the surface electron states.
The "numerically exact" calculations show that, in the nanowire, the surface  wave-function density probability exhibits a significant increase at the corners.   
Actually, in the upper right panel of Fig. \ref{fig:8}, in the case of the maximum of the gap ($L_{min}=90 \AA$, for aspect ratio 1/6), the spatial probability density of the first subband state is higher at the corners. Actually, it shows some analogies with the state corresponding to the maximum of the gap in the cylindrical nanowire (see Fig. \ref{fig:3}).  In fact, as in the cylinder, it is not strongly confined on the perimeter. However, it shows also an enhanced corner density which represents the fingerprint of the rectangular cross-sections. It is worthy noticing that the formation of corner states is a consequence of the matching between the wave-function along $x$ and $y$ sides.
On the other hand, in the upper left panel of Fig. \ref{fig:8}, the spatial probability density of the first subband in the case of the minimum of the gap ($L_{min}=60 \AA$, for aspect ratio 1/6) is completely different. A sort of side state takes place. It is clear that, at the minimum of the gap, one gets a complete decoupling of the two sides with smaller length along the perimeter. In this way the system reproduces  the case of TI with a single confinement in the direction orthogonal to shorter sides. Therefore, one expects that the Dirac cone is recovered (zero gap in the upper panel of Fig. \ref{fig:7}).  

For all the aspect ratios, as shown in the lower  panel of Fig. \ref{fig:7}, the gap vanishes in the presence of appropriate magnetic fields. Asymptotically, for all the geometries, the energy gap closes at half quantum flux. Again the case of ratio 1/6 is the most interesting. Indeed, as expected, for the sizes $L_{min}$ corresponding to the zero of the gap in the absence of magnetic field, the magnetic flux for gap closing is very small.   

Finally, we focus on the spin-texture corresponding to the first subband state.  We follow the same procedure used in the case of the cylinder, diagonalizing the spin operators in the two-dimensional subspace corresponding at fixed energy.  Considering the confinement in the xy plane with invariance along z, the spin behaves in the same way as in the cylinder, i.e. it rotates around the perimeter (see the bottom right panel of Fig. \ref{fig:8} corresponding to $L_{min}=90 \AA$ for aspect ratio 1/6). However, for $L_{min}$ at the minimum of the gap  (see the bottom left panel of Fig. \ref{fig:8} corresponding to $L_{min}=60 \AA$ for aspect ratio 1/6), there is a decoupling of the spin texture along the shortest sides confirming that the system forms only two side states  without any presence of corner states.

The case of confinement in the xz plane and invariance along the y direction is more complicated.  If we consider a square cross-section of small size, all the three eigenvalues $\lambda_x$,  $\lambda_y $ and $\lambda_z$ of the spin components are not zero and comparable. In this way, the eigenvalues of the spin are such that, along the x direction of the cross-section of the wire, there is $\lambda_x$ on one side and $-\lambda_x$ on the other.  Instead, the value of $\lambda_y$ and $\lambda_z $ is different from zero on all four sides of the square cross-section. If we increase the size of the square, however, we observe that the two components $\lambda_y$ and $\lambda_z$ decrease slowly, remaining only the component $\lambda_x$ unchanged.  In the asymptotic limit, for the case of translational invariance along y direction, the spin still turns around the perimeter, but in such a way that it vanishes along the two sides in the z direction. This occurs more easily in a geometry with a strong asymmetry along the sides, for example  when one decreases the aspect ratios. The behavior of the spin for this confinement can be ascribed to the fact that, already with a single in-plane confinement, for example a confinement of the three-dimensional material only along the x direction,  the corresponding surface states along $yz$ plane are not described by a Rashba hamiltonian responsible for a simple spin-momentum locking \cite{Sen}. In particular, it is along the z direction that the spin behavior is more complicated. Indeed, in the case of nanowire with a double confinement, it is on the z side that the spin has a different behavior in comparison with that of the cylinder and rectangular wire with translational axes along z direction (confinement in xy plane).

\section{Conclusions and discussions}

In this paper, we have studied a simple model for the electronic structure of a $Bi_2Se_3$ nanowire with different geometries of the cross-section. In particular, we have focused on the role played by the Berry phase in opening a gap at the Dirac cone for finite area cross-sections. Moreover, we have analyzed the effects of  a magnetic field applied along the translational axis of the nanowire that is able to close the energy gap. Therefore, we have pointed out the relevance of quantum interference effects on measurable quantities of the nanowire, such as the gap and the magnetic flux responsible for the gap closure.   

First, we have analyzed the electronic states of a cylindrical nanowire. We have improved the approximate analytical solution proposed in Ref. [\onlinecite{Imura}] extending its validity to lower radii $R$ of the cylindrical nanowire. This has allowed to show that  the closure of the gap takes place for magnetic fields which are larger than those corresponding to half quantum flux and which increase with decreasing the nanowire radius.  An exact approach based on numerical diagonalizations has been implemented not only to check the accuracy of the analytical solutions but also to study cylinders with small radii. With decreasing the radius, the numerical approach confirms the presence of a maximum and a minimum of the gap energy as a function of the radius in semiquantitative agreement with the analytical solution.

Then, we have numerically studied the electronic states of a nanowire with rectangular cross-sections focusing on different aspect ratios between the side lengths of the rectangle.  
In particular, we have analyzed the properties of the edge states finding that they form without any additional corner potential as suggested in the literature \cite{Sen}.
Moreover, we have pointed out that the gap size and the related behavior in the presence of  a magnetic field  are qualitatively similar but they quantitatively depend on the size and geometry of the nanowire. All the methods used in this paper have confirmed the presence of a minimum and maximum of the energy gap at the $\Gamma$ point in the range of characteristic cross-section lengths from $50$ to $100 \AA$ and the resulting variation of the magnetic flux for the gap closing close to the minimum. One of the most important results is obtained in the case of a rectangle with aspect ratio equal to $1/6$. Indeed, the minimum coincides nearly with zero energy implying that the gap can be closed without any magnetic field when the length $L_{min}$ of the smaller side  is the order of $60 \AA$.  Therefore, a decoupling of the system takes place on the two sides with the shortest length along the perimeter, in strong analogy with  the case of a single confinement in the direction orthogonal to these sides. Close to the minimum of the gap, the first sub-band state shows  both a spatial probability density and a spin polarization concentrated almost along the shortest sides of the perimeter. 

In this paper, interesting electronic properties of the nanowire have been obtained in the range of cross section characteristic lengths from $50$ to $100 \AA$, where we expect that effects due to quantum confinements of the wave-function are not relevant.   Actually, we have considered energy levels of the first sub-bands well within the gap of the three dimensional bulk. Only for cross section characteristic lengths smaller than $10 \AA$, the first sub-band state has energies comparable with the bands of the three dimensional bulk. Indeed, it would be very interesting to theoretically analyze the transport properties of the wire \cite{nuovo1,nuovo} by decreasing the cross-section area and varying the chemical  potential. Moreover,  the thermoelectric properties in TI \cite{Xu} would be enhanced in a nanowire of Bismuth based materials which are known for their large Seebeck effects and thermoelectric figure of merit \cite{nuovo2}.  These studies would naturally require the introduction of disorder potentials inevitably present in these materials.  Work in this direction is in progress.

\begin{acknowledgments}
C. A. P. acknowledges partial financial support from the Progetto Premiale CNR/INFN EOS \textit{ Organic Electronics for Innovative Research Instrumentation}. C. A. P. and V. C. acknowledge partial financial support from the regione Campania project L.R. N.5/2007  \textit{ Role of interfaces in magnetic strongly correlated oxides: manganite heterostructures.} 
\end{acknowledgments}

\begin{appendix}

\section{Numerical approach}
In this Appendix, we provide details about the numerical approach used in this paper focusing on the case of the cilindrical nanowire.  

\begin{figure}[h!]
\centering
\includegraphics[width=8.cm]{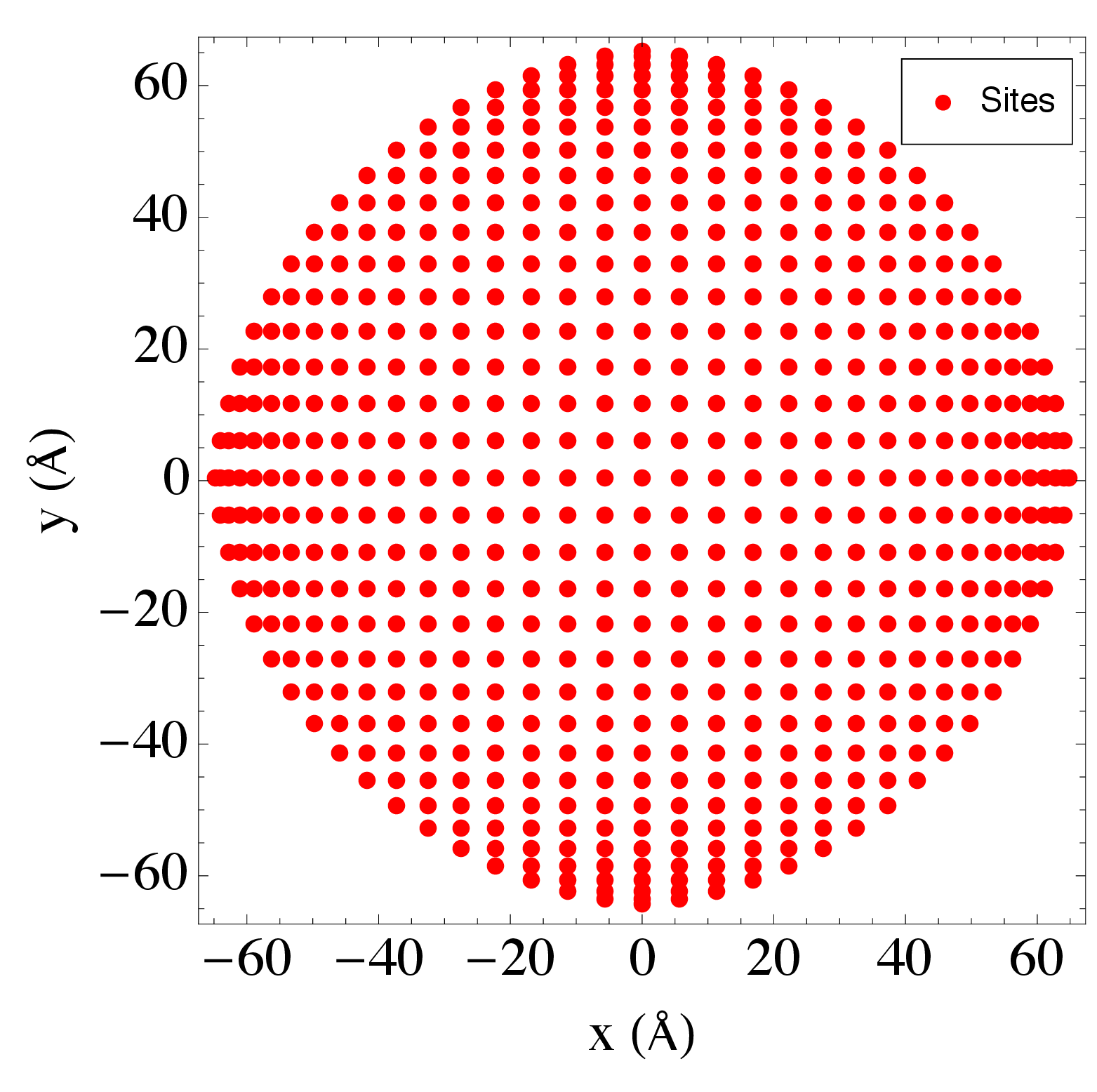}
\label{fig:6}
\caption{ Site positions (in the real space) with variable mesh in the case of radius $R = 65 \AA$. A perfect circumference on the boundary is numerically simulated.}
\end{figure}

The numerical approach exactly solves the eigenvalue/eigenstate problem. It is based on a discretization of the Hamiltonian in eq.(\ref{eq:1}), reducing the continuum to a discrete lattice. For square/rectangular nanowires, one considers square/rectangular lattices with fixed lattice parameters. In the case of the cylindrical nanowire, the procedure is more complicated since one has to discretize the continuum model with a circumference as boundary. This issue is particularly relevant since we want to study the surface states. To this aim, we choose a mesh such that the sites present in bulk are more rarefied and those on the edge are more dense, as shown in Fig. \ref{fig:6}. In this way, we are able to create a perfect circumference, simulating very well the surface. Using a numerical diagonalization procedure with open boundary conditions, one gets  bulk and surface states making a comparison with the analytical solutions. The numerical approach perfectly reproduces the analytical results for large radii, where the analytical solutions are valid.  Therefore, the numerical approach allows to make an extrapolation to the continuum limit. Since we have considered a variable mesh, we make the limit of the average lattice parameter of the lattice to zero.

To derive the matrix elements of the Hamiltonian, we exploit the Taylor expansion of the first and second derivatives present in Eq.(\ref{eq:1}). If we consider the i-th site, where the index $i=x,y$, we denote $a_i$ its distance from the nearest neighbor on the left,   $a_{i+1}$  its distance from the nearest neighbor on the right.  Based on these notations, we write the first derivative of the wave function $\psi$ in terms of finite differences:

\begin{eqnarray}
\label{eq:30}
	\psi_{i+1}&=&\psi_i - \psi^{'}_i a_i \nonumber\\
	\psi_{i-1}&=&\psi_i + \psi^{'}_i a_{i+1} .
\end{eqnarray}
 
Doing the same with the expansion to the second order of the eq.(\ref{eq:30}), one gets

Therefore, we are able to write the Hamiltonian (\ref{eq:1}) as a matrix $(4N_xN_y)^2$, where $N_x$ and $N_y$ are the site numbers along x and y direction in the non-uniform grid.

\begin{eqnarray}
\psi_i^{'}  & = &\frac{\psi_{i-1} - \psi_{i+1}}{a_i+a_{i+1}} \nonumber\\
\psi_i^{''}  & =& \frac{2\psi_{i+1}}{a_i(a_{i+1}+a_i)}+\frac{2\psi_{i-1}} {a_{i+1}(a_{i+1}+a_i)} -\frac{2\psi_{i,j}}{a_ia_{i+1}}
\label{eq:33}
\end{eqnarray}

This grid is constructed such that each distance scales as a function of sine and cosine. Hence, it is possible to select points that are located at a distance from the center less or equal to half side of the square which contains the  circle (see Fig. \ref{fig:6}). The elements which are outside the circle must be put to zero. Therefore, the matrix will contain a number of rows and columns containing zeros. 
Once allocated the matrix, it is possible to obtain eigenvalues and eigenvectors with any  diagonalization routine. 

Once found eigenstates and eigenvalues, we can get the orientation of spin on the surface building up a matrix similar to that in equation (\ref{eq:10}) of section III. This time, however, the representation is made considering a discrete lattice and not a continuum.  Therefore, considering for simplicity $k_z=0$ and indicating with 0 and $-1$ the two eigenvectors corresponding to the same eigenvalue, one gets the following matrix 
for the spin components at each lattice site:

 \begin{equation}
\begin{pmatrix} 
\label{eq:34}
  \langle \psi_0(i) |S_j|\psi_0(i)  \rangle      &      \langle \psi_{0}(i) |S_j|\psi_{-1}(i)  \rangle\nonumber  \\ 
  \langle \psi_{-1}(i)|S_j |\psi_0(i) \rangle   &      \langle \psi_{-1}(i) |S_j|\psi_{-1}(i)   \rangle \\
  \end{pmatrix},
\end{equation}
with $j=x,y,z$. Therefore, the couple $(r,\phi)$ of the wave function $\psi_m(r,\phi)$ in Eq.(\ref{eq:10}) is replaced by the index $i$ that labels the sites as in Fig. \ref{fig:6}. The eigenvalues of the equation (\ref{eq:34}) for each site $i$ will provide the spin orientation as found in bottom panel of Fig. \ref{fig:4} in section III.

\end{appendix}


\end{document}